\documentclass[preprint2]{aastex}
\usepackage{rotating}
\usepackage{graphicx}
\usepackage{grffile}
\usepackage{multicol}
\usepackage{longtable}
\usepackage{pdflscape}

\begin{document}
\setcounter{secnumdepth}{2}
\title{The TRENDS High-Contrast Imaging Survey. VI. \\ Discovery of a Mass, Age, and Metallicity Benchmark Brown Dwarf}
\author{Justin R. Crepp\altaffilmark{1}, Erica J. Gonzales\altaffilmark{1}, Eric B. Bechter\altaffilmark{1}, Benjamin T. Montet\altaffilmark{2,3}, John Asher Johnson\altaffilmark{2}, Danielle Piskorz\altaffilmark{3}, Andrew W. Howard\altaffilmark{4}, Howard Isaacson\altaffilmark{5}}
\email{jcrepp@nd.edu} 
\altaffiltext{1}{Department of Physics, University of Notre Dame, 225 Nieuwland Science Hall, Notre Dame, IN, 46556, USA}
\altaffiltext{2}{Harvard-Smithsonian Center for Astrophysics, 60 Garden Street, Cambridge, MA 02138, USA}
\altaffiltext{3}{Division of Geological and Planetary Sciences, California Institute of Technology, 1200 E California Blvd., Pasadena, CA 91125}
\altaffiltext{4}{Institute for Astronomy, University of Hawaii, 2680 Woodlawn Drive, Honolulu, HI 96822}
\altaffiltext{5}{Department of Astronomy, University of California, Berkeley, CA 94720} 
\date{}

\begin{abstract}
The mass and age of substellar objects are degenerate parameters leaving the evolutionary state of brown dwarfs ambiguous without additional information. Theoretical models are normally used to help distinguish between old, massive brown dwarfs and young, low mass brown dwarfs but these models have yet to be properly calibrated. We have carried out an infrared high-contrast imaging program with the goal of detecting substellar objects as companions to nearby stars to help break degeneracies in inferred physical properties such as mass, age, and composition. Rather than using imaging observations alone, our targets are pre-selected based on the existence of dynamical accelerations informed from years of stellar radial velocity (RV) measurements. In this paper, we present the discovery of a rare benchmark brown dwarf orbiting the nearby ($d=18.69\pm0.19$ pc), solar-type (G9V) star HD~4747 ([Fe/H]=$-0.22\pm0.04$) with a projected separation of only $\rho=11.3\pm0.2$ AU ($\theta=0.6''$). Precise Doppler measurements taken over 18 years reveal the companion's orbit and allow us to place strong constraints on its mass using dynamics ($m \sin i = 55.3\pm1.9M_{\rm Jup}$). Relative photometry ($\Delta K_s=9.05\pm0.14$, $M_{K_s}=13.00\pm0.14$, $K_s - L' = 1.34\pm0.46$) indicates that HD~4747~B is most-likely a late-type L-dwarf and, if near the L/T transition, an intriguing source for studying cloud physics, variability, and polarization. We estimate a model-dependent mass of $m=72^{+3}_{-13}M_{\rm Jup}$ for an age of $3.3^{+2.3}_{-1.9}$ Gyr based on gyrochronology. Combining astrometric measurements with RV data, we calculate the companion dynamical mass ($m=60.2\pm3.3M_{\rm Jup}$) and orbit ($e=0.740\pm0.002$) directly. As a new mass, age, and metallicity benchmark, HD~4747~B will serve as a laboratory for precision astrophysics to test theoretical models that describe the emergent radiation of brown dwarfs. 
\end{abstract}

\section{INTRODUCTION}\label{sec:intro}
Mass governs the evolution of substellar objects as it determines the path or ``track" that their luminosity and effective temperature follow in time \citep{stevenson_91}. Having insufficient pressure at their cores to burn hydrogen through sustained nuclear fusion, brown dwarfs gradually emit less light as they cool with time \citep{baraffe_03}. Although the relative brightness of a brown dwarf may be measured in various filters, incident flux from a given distance does not uniquely determine a brown dwarf's mass because mass and age are degenerate parameters. Additional information is required.   

The first brown dwarfs were discovered using imaging \citep{rebolo_95,becklin_88,nakajima_95}. However, even with direct spectroscopic follow-up observations the mass of most brown dwarfs is still poorly constrained. Mass estimates are instead nominally based on theoretical evolutionary models, which have yet to be properly calibrated \citep{dupuy_14}. Further, model-dependent mass estimates rely on a precise age determination, a notoriously difficult physical property to ascertain \citep{soderblom_10}. As a result, substellar objects that are not members of binary or higher-order multiple systems often have highly uncertain masses \citep{dupuy_kraus_13}. In fact, some directly imaged brown dwarfs have been initially mistaken for extrasolar planets due to the above ambiguity \citep{carson_13,hinkley_13}. 

In this paper, we present the direct detection of a faint co-moving companion orbiting the nearby ($d=18.7$ pc) star HD~4747 (Table 1). HD~4747 is a bright ($V=7.16$) G9V star originally targeted with precise Doppler measurements in 1996 using the High Resolution Echelle Spectrometer (HIRES) at Keck \citep{vogt_94}. Using five years of data, a strong acceleration was noticed and attributed to an orbiting companion \citep{nidever_02}. These initial measurements caught the companion near periastron and now provide important constraints on the orbital elements but, at the time, were insufficient to determine a unique period or velocity semi-amplitude due to the limited time baseline.  

\begin{table}[!ht]
\begin{tabular}{lc}
\hline
\multicolumn{2}{c}{HD~4747 Properties}     \\
\hline
right ascension [J2000]            &   00 49 26.77      \\
declination [J2000]                   &    -23$^{\circ}$ 12 44.93        \\
$B$                                           &   7.92     \\
$V$                                           &   7.16     \\
$R$                                           &   6.73     \\
$I$                                            &    6.34     \\
$J_{\tiny{\rm 2MASS}}$            &    $5.813\pm0.021$  \\
$H_{\tiny{\rm 2MASS}}$            &   $5.433\pm0.049$ \\
$K_s$                                        &   $5.305\pm0.029$  \\
$L'$                                            &    $5.2\pm0.1$         \\
d [pc]                                          &    $18.69\pm0.19$   \\
proper motion [mas/yr]              &    $516.92\pm0.55$  E \\
                                                   &   $120.05\pm0.45$ N  \\
\hline
Mass [$M_{\odot}$]           &    $0.82\pm0.04$        \\
Radius [$R_{\odot}$]         &     $0.79\pm0.03$       \\
Age [Gyr]                            &     $3.3^{+2.3}_{-1.9}$   \\
$\mbox{[Fe/H]}$                 &      $-0.22\pm0.04$      \\
log g [cm $\mbox{s}^{-2}$]  &  $4.65\pm0.06$       \\
$T_{\rm eff}$ [$K$]              &    $5340\pm40$        \\
Spectral Type                      &       G9V        \\
v sini   [km/s]                        &   $1.1\pm0.5$        \\
\hline
\end{tabular}
\caption{Observational and physical properties of the HD~4747 system from \citealt{montes_01} (spectral type), \citealt{van_leeuwen_07} (parallax), \citealt{koen_10} (visible photometry), \citealt{cutri_03} (NIR photometry) and the Spectroscopic Properties of Cool Stars (SPOCS) database \citep{valenti_fischer_05}. The L' magnitude is estimated from a black-body fit to the stellar spectral energy distribution.} 
\label{tab:starprops}
\end{table} 

\begin{figure*}[!t]
\includegraphics[height=4.8in]{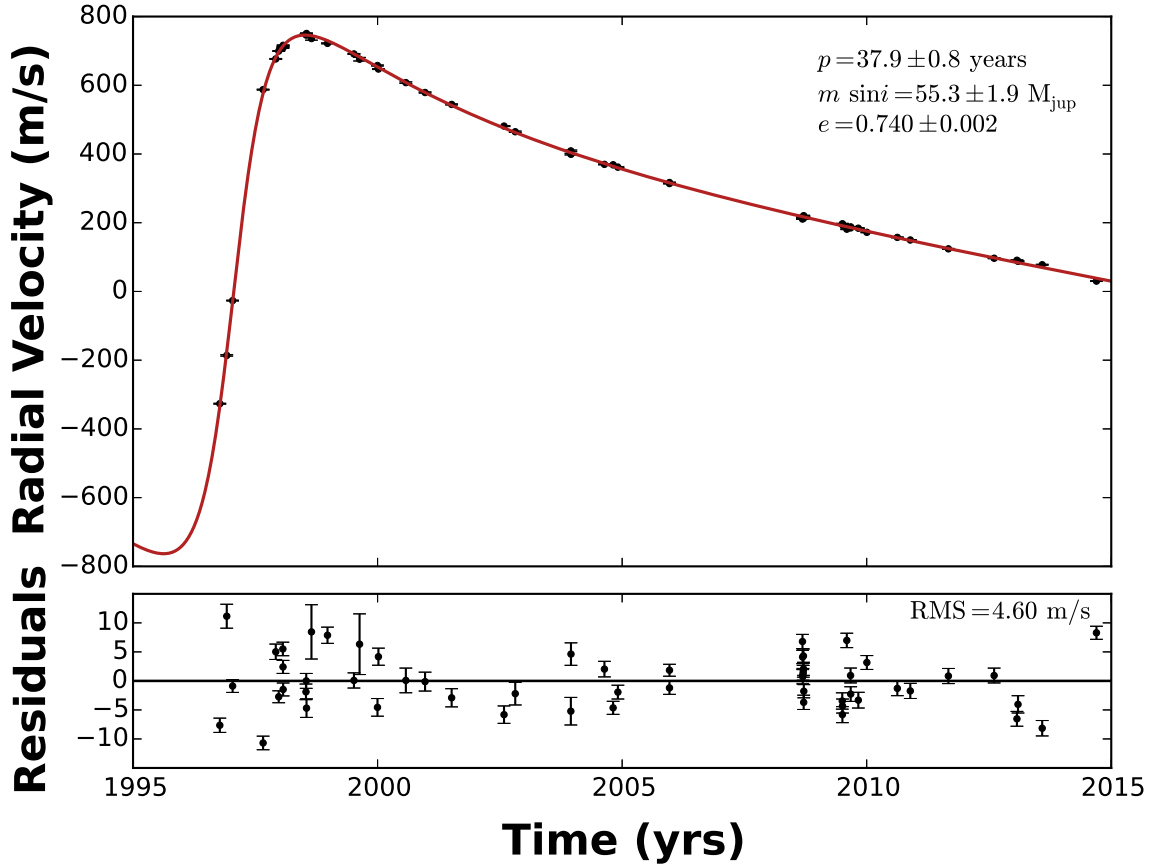} 
\caption{Doppler radial velocity measurements of HD~4747. A changing acceleration is consistent with a Keplerian orbit indicating a companion with orbital period amenable to direct imaging. We used this data to infer the approximate angular separation of HD~4747~B in advance of high-contrast AO observations that detected the companion directly.} 
\label{fig:rvs}
\end{figure*}  

HD~4747 was subsequently identified by \citealt{raghavan_10} as an ``accelerating proper-motion" star based on comparison of the time resolved system proper-motion as tabulated in the \emph{Tycho-2} catalog from \emph{Hipparcos}. In a search for brown dwarf companions, \citet{sahlmann_11} were able to corroborate the single-line spectroscopic binary nature of HD~4747 using the Coralie and HARPS spectrographs. In addition to Doppler observations, \citet{sahlmann_11} investigated the \emph{Hipparcos} astrometry measurements of the host star in an attempt to constrain the companion inclination and mass but found that the orbital phase coverage, which spanned from 1989-1993, was too low to justify a joint analysis. Using RV's alone, they found a best-fit solution for the orbital period of $P=31.8^{+3.0}_{-3.1}$ years with an eccentricity of $e=0.723^{+0.013}_{-0.013}$. We show below using more recent Doppler measurements and direct imaging that their period is only slightly underestimated and the eccentricity is correct to within uncertainties.  

The California Planet Search (CPS; \citealt{howard_10a}) has now measured precise RVs of HD 4747 for nearly two decades.  This star is also one of 166 G and K dwarfs within 25 pc with intensive CPS observations in the Eta-Earth Survey \citep{howard_10b}. Table 1 lists the host star physical properties based on (non-iodine) template spectra taken in concert with Doppler measurements. Follow-up spectra were acquired with HIRES and analyzed using standard techniques that utilize an iodine vapor cell for instrument calibration to record precise RV measurements \citep{howard_13}. Upon being linked with the original \citealt{nidever_02} data set, the full RV time series measurements now span a total of 18 years (Fig. 1). 

More recently, HD~4747 was flagged as a high-priority target for the TRENDS high-contrast imaging program based on the fact that the companion must be close to apoastron (Crepp et al. 2012b, 2013a; 2013b; 2014; Montet et al. 2014). Based on the available RV information, which reveals significant curvature including an inflection point, we were able to deduce the companion approximate angular separation well in advance of our adaptive optics (AO) imaging observations (see $\S$\ref{sec:imaging}). This paper presents the results of our high-contrast imaging observations using the NIRC2 instrument at Keck that identify the companion responsible for producing the Doppler acceleration.  

\begin{figure*}[!t]
\begin{center}
\includegraphics[height=3.0in]{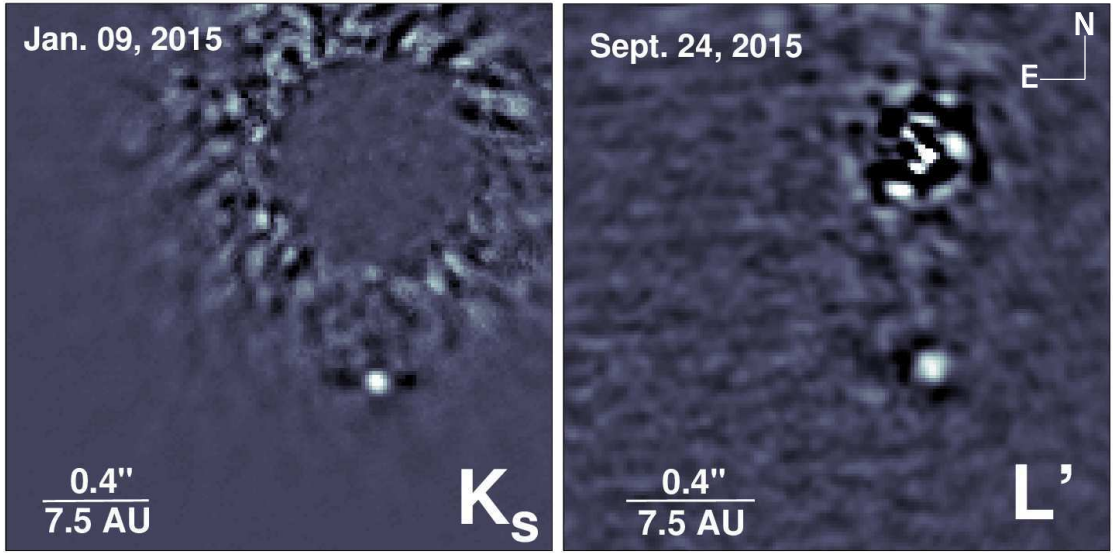} 
\caption{Confirmation images of HD~4747~B taken with NIRC2 in the $K_s$-band on 2015-January-09 (left) and $L'$-band on 2015-September-24 (right). The scale and orientation of the images is identical. The measured angular separation of $\rho=0.6''$ corresponds to a projected separation of only 11.3 AU, placing the companion squarely in the brown dwarf desert.}
\end{center}
\end{figure*}  


 
\section{IMAGING DISCOVERY}\label{sec:imaging}
HD~4747 was initially observed with NIRC2 at Keck on August 18, 2013. A quick snap-shot AO image (90 second integration time) taken without the coronagraph using the $K_{\rm cont}$ filter resulted in a non-detection, thus effectively eliminating the possibility of the companion being a face-on binary. HD~4747 was reobserved using NIRC2 on 2014 October 12. This time, a wider filter was used ($K_s$) and the bright host star was placed behind the 600 mas wide coronagraphic mask to block on-axis light and allow for longer integration times. The image derotator was turned off to enable point-spread-function (PSF) subtraction of scattered starlight through angular differential imaging \citep{marois_08}. Additional calibration images were taken to measure the flux ratio of the companion relative to the star. 

\begin{table}[!t]
\centerline{
\begin{tabular}{lcccccc}
\hline
\hline
  Date [UT]          &     Filter    &     $\Delta t$ [m] &   Airmass   &  $\Delta \pi$ [$^{\circ}$]   \\
\hline
\hline        
2014-10-12         &     $K_s$     &  94   &   1.38-1.52     &  41.6  \\
2015-01-09         &     $K_s$     &  39   &   1.44-1.61    &   13.9 \\
2015-09-24         &      $L'$       &  40   &    1.36-1.43    &   28.8 \\
\hline
\hline
\end{tabular}}
\caption{Summary of high-contrast imaging measurements taken with NIRC2 including on-source integration time ($\Delta t$) in minutes, airmass range over the course of observation, and change in parallactic angle ($\Delta \pi$).}
\label{tab:obs}
\end{table}

\begin{table*}[!t]
\centerline{
\begin{tabular}{lccccc}
\hline
\hline
  Date [UT]      & JD-2,450,000         &     Filter    &      $\rho$ [mas]      &    Position Angle [$^{\circ}$]     &   Proj. Sep. [AU]   \\
\hline
\hline        
2014-10-12           & 6,942.8   &     $K_s$     &     $606.5\pm7.0$    &    $180.04^{\circ}\pm0.62^{\circ}$       &    $11.33\pm0.17$ \\
2015-01-09           & 7,031.7    &     $K_s$    &     $606.6\pm6.4$    &    $180.52^{\circ}\pm0.58^{\circ}$       &    $11.34\pm0.16$ \\
2015-09-24           & 7,289.9    &      $L'$      &     $604\pm7$           &     $184.9^{\circ}\pm0.9^{\circ}$         &    $11.3\pm0.2$      \\ 
\hline
\hline
\end{tabular}}
\caption{Summary of astrometric measurements using NIRC2.}
\label{tab:astrometry}
\end{table*}

Upon detecting an off-axis candidate source from the 2014 October data set, we acquired follow-up observations using NIRC2 on 2015 January 09. The observations were again recorded in $K_s$ to assess whether the candidate companion was co-moving with its host star. We were unable to acquire ADI measurements in complementary filters as HD~4747 was low on the horizon when observed from Mauna Kea. The following observing season (2015 September 24), we obtained $L'$ data with NIRC2 to assess the companion infrared spectral energy distribution and help to infer a temperature. In each case, the companion was unambiguously recovered with a high signal-to-noise ratio (Fig.~2). Table~\ref{tab:obs} summarizes our measurements. 

The NIRC2 data were analyzed using standard high-contrast imaging techniques as described in \citealt{crepp_13b} and other TRENDS articles. We note that the companion cannot be seen in raw frames since it is fainter than stellar speckles. As customary, we inject fake companions with similar brightness as the detected source using an iterative method to calibrate the PSF subtraction procedure in order to account for unwanted flux removal (self-subtraction) which can affect both astrometry and photometry \citep{crepp_11}. 

\begin{figure}[!t]
\includegraphics[height=2.82in]{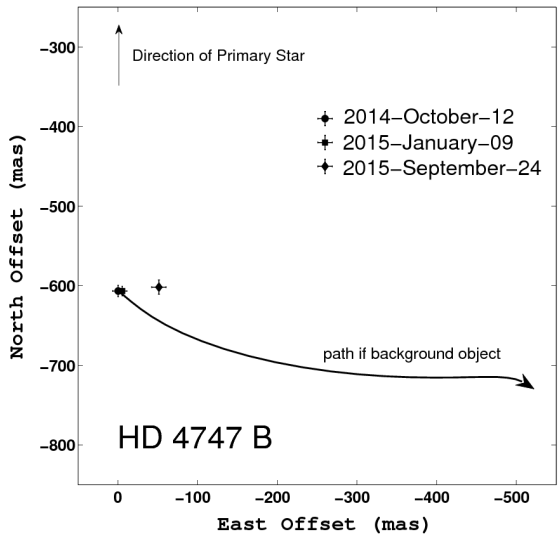} 
\caption{Astrometric confirmation plot showing three successive epochs of HD~4747~B (represented by circle, square, diamond) compared to the path of a distant background object. The companion is unambiguously associated with HD~4747~A, which is located at the origin of the plot and indicated by an arrow pointing north.}
\label{fig:confirm}
\end{figure}

HD~4747~B has a projected separation of $\theta=0.6$". Given that the primary star has such a large proper-motion (Table 1), we were able to confirm association with only three months of imaging time-baseline between the first and second imaging epochs. A third data point, the L-band measurement, reveals orbital motion in a counter-clockwise direction. Table~\ref{tab:astrometry} summarizes our astrometry measurements, and Fig.~\ref{fig:confirm} shows the relative motion of HD~4747~B compared to the null hypothesis of a distant background object demonstrating association. 

\section{CHARACTERIZATION}

\begin{figure*}[!t]
\includegraphics[height=4.0in]{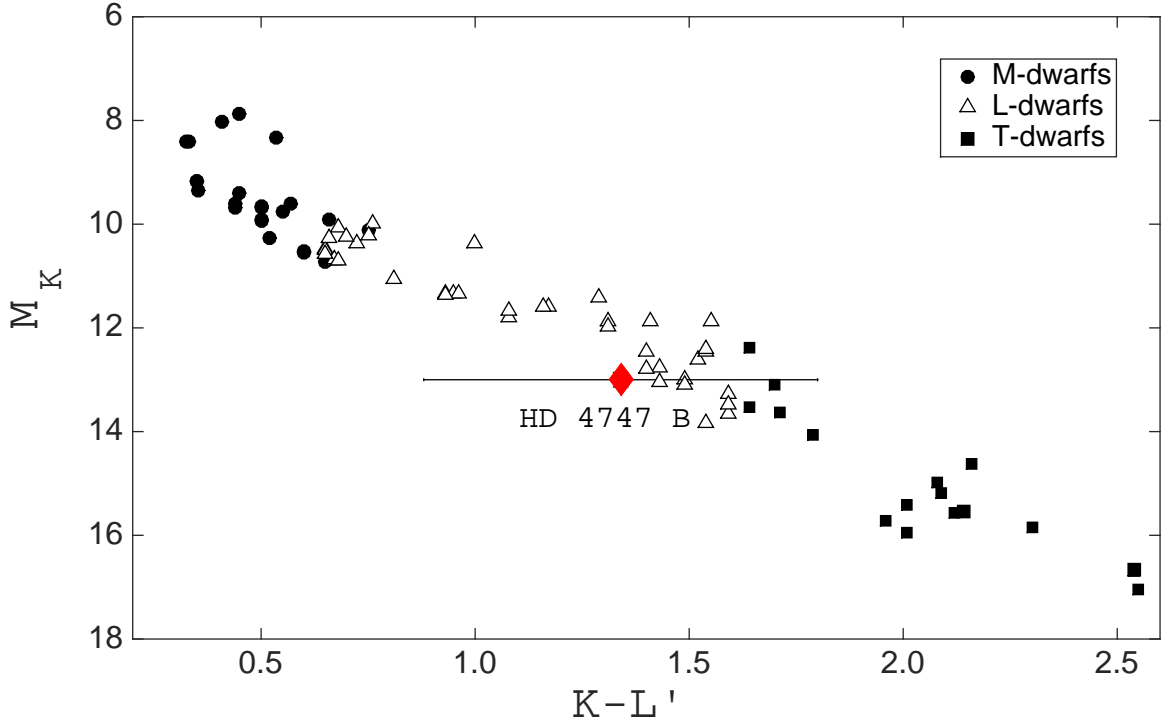} 
\caption{Color-magnitude diagram comparing HD~4747~B to cold dwarfs from \citet{leggett_02} and \citet{dupuy_12}. HD~4747~B (red diamond) is most likely a mid-to-late-type L-dwarf. It may be close to the L/T transition.}
\label{fig:CMD}
\end{figure*}

We find that HD~4747~B is $\Delta K_s = 9.05\pm0.14$ mags fainter than its parent star corresponding to an absolute magnitude of $M_{K_s}=13.00\pm0.14$ (Table~\ref{tab:phot}). The system parallax is known to within 1\% and therefore does not contribute appreciably to this uncertainty. Similarly, reducing the NIRC L'-band data, we find $M_{L'}=11.66\pm0.44$ resulting in an infrared color of $K_s-L'=1.34\pm0.46$. The $K_s-L'$ color is useful because it correlates monotonically with spectral type in the LT-dwarf range \citep{leggett_02,skemer_16}. Figure~\ref{fig:CMD} shows a color-magnitude diagram comparing HD~4747~B to other cold dwarfs. Our photometry indirectly suggests that the companion is a late-type L-dwarf and perhaps near the L/T transition. We do not attempt to assign a spectral type without spectroscopy. A forth-coming paper will address this topic. For completeness, we report measured flux values of $\Delta K_s=8.98\pm0.18$ from the October 2014 data set and $\Delta K_s=9.15\pm0.22$ from the January 2015 data set. Table~\ref{tab:phot} shows the weighted average. We might expect small amplitude K-band variability from an L-dwarf from evolving cloud coverage though our observations are not sufficiently precise to study these effects \citep{metchev_15}. If HD~4747~B is indeed near the L/T transition, it would serve as an interesting brown dwarf to study differential opacities, methane absorption, variability, and polarization (e.g. \citealt{jensen-clem_16}). 

We estimate a model-dependent companion mass using our $M_{K_s}$ value, which has a much smaller uncertainty than the $M_{L'}$ absolute magnitude (Table~\ref{tab:phot}). Given the host star $B-V=0.76$ color, estimated rotational spin period ($P_{\rm rot}\sim27$ days), and gyro-chronology relations from \citet{mamajek_hillenbrand_08}, we estimate a system age of $3.3^{+2.3}_{-1.9}$ Gyr taking into account scatter in the empirical relation. Interpolating the theoretical cooling curves from \citet{baraffe_03}, we find a model-dependent mass of $m_{\rm model}=72^{+3}_{-13}M_{\rm Jup}$ and $T_{\rm eff}\approx 1700\pm100$ K, assuming co-evolution between the companion and host star. The large asymmetry in mass uncertainty stems from the fact that the companion straddles the brown dwarf / stellar boundary given the uncertainty in its age. 

In summary, HD~4747~B has an intrinsic brightness, color, and model-dependent mass that are self-consistent. In the next section, we perform a comprehensive dynamical analysis that jointly fits the long-term RV measurements with astrometry to understand the companion mass and orbit in more detail. We show that the physical separation and companion brightness are also consistent with producing the measured RV acceleration. 

\begin{table}[!ht]
\centerline{
\begin{tabular}{lc}
\hline
\hline
\multicolumn{2}{c}{HD~4747~B Photometry}     \\
\hline
\hline
$\Delta K_s$                 &      $9.05\pm0.14$       \\
$\Delta L'$                     &     $7.82\pm0.43$       \\
$K_s$                            &     $14.36\pm0.14$     \\
$L'$                               &     $13.02\pm0.44$      \\
$M_{K_s}$                    &     $13.00\pm0.14$      \\
$M_{L'}$                       &      $11.66\pm0.44$      \\
$K_s-L'$                       &       $1.34\pm0.46$       \\
$T_{\rm eff}  [K]$          &      $\approx1700\pm100$    \\
Spec. Type                   &       $\approx L8$              \\
\hline
\hline
\end{tabular}}
\caption{Companion magnitudes, infrared color, and estimated effective temperature from photometry using the \citet{baraffe_03} thermal evolutionary models.}
\label{tab:phot}
\end{table} 

\begin{table}[!t]
\centerline{
\begin{tabular}{lc}
\hline
\hline
\multicolumn{2}{c}{Orbital Fitting Results} \\
\hline
\hline
Parameter                        &     Value (68\%)       \\   
\hline      
K  [m/s]                             &  $755.3^{+12.4}_{-11.6}$        \\                            
P [years]                           &  $37.88^{+0.86}_{-0.78}$        \\                     
e                                       &  $0.740^{+0.002}_{-0.002}$    \\              
$\omega$ [$^{\circ}$]       &  $269.1^{+0.6}_{-0.5}$             \\   
$t_p$ [year]                      &  $1997.04^{+0.02}_{-0.02}$    \\       
$i$ [$^{\circ}$]                  &  $66.8^{+4.7}_{-6.4}$               \\   
$\Omega$ [$^{\circ}$]      &  $188.3^{+3.2}_{-2.6}$             \\   
$a$ [AU]                           &  $16.4^{+3.9}_{-3.3}$               \\
\hline
Global offset [m/s]             & 	$-214.9^{+10.9}_{-11.6}$  \\
Diff. offset [m/s]      	          &   $13.5^{+2.8}_{-2.7}$         \\
Jitter [m/s]                 	     &   $4.0^{+0.9}_{-0.7}$           \\
\hline
\hline
\end{tabular}}
\caption{Orbital parameters for HD~4747~B from a joint-fit to the imaging and RV data. The best-fitting stellar jitter value is consistent with RV residuals shown in Fig.~\ref{fig:rvs}.}
\label{tab:orbitfit}
\end{table}

\section{ORBIT AND DYNAMICAL MASS}
We constrain the mass and orbit of HD~4747~B using all available RV observations along with three direct imaging measurements. The RV time series spans 18 years (47\% of the complete orbit) while the astrometry spans 0.95 years. Each measurement achieves a high signal-to-noise ratio ($\sim100$), i.e. the uncertainty in individual measurements is small compared to the Doppler semi-amplitude and projected angular separation on the sky. This enables convergence of all orbit parameters to a unique solution. In particular, the RV inflection points shown in Fig.~\ref{fig:rvs} constrain the velocity semi-amplitude ($K$), eccentricity ($e$), argument of periastron ($\omega$), time of periastron passage ($t_p$), and orbital period ($P$). Only several astrometric measurements are required to constrain the orbit semi-major axis ($a$), inclination ($i$), and longitude of the ascending node ($\Omega$) given the system parallax and close orbital separation of $\rho=11.3$ AU. The complete orbit in turn reveals the system total mass through Kepler's equation, as well as the companion mass through $m \sin{i}$ and the orbit inclination.

We perform a joint fit on the combined data set using Bayesian inference. Measurements from Table~\ref{tab:astrometry} and Table~\ref{tab:rvtable} are analyzed with a Markov Chain Monte Carlo (MCMC) simulation. Our techniques follow closely that of \citealt{crepp_12a}. However, instead of using the Metropolis-Hastings algorithm to explore the orbital parameter distributions we employ an affine (aspect-ratio) invariant ensemble sampler. Originally proposed by \citealt{goodman_10} but written in python and made publicly available by \citep{foreman-mackey_13}, this MCMC code (``emcee: The MCMC Hammer") runs on multiple CPU's, efficiently handles correlations between fit parameters, which greatly improves convergence, and enables model fitting in a high dimensional space as is needed to solve Kepler's problem. 

The results of our orbital fitting analysis are summarized in Table~\ref{tab:orbitfit}. We find that HD~4747~B has an orbital period ($P=37.88^{+0.86}_{-0.78}$ years) and semimajor axis ($a=16.4^{+3.9}_{-3.3}$) that places it within the brown dwarf ``desert," a range of orbital radii in close proximity to Sun-like stars where very few substellar companions form compared to objects found in the field \citep{liu_02}. We also corroborate the results of \citet{sahlmann_11} who found HD~4747~B to have a very high eccentricity, a feature than can readily be seen from Fig.~\ref{fig:rvs}. 

\begin{figure}[!t]
\includegraphics[height=2.5in]{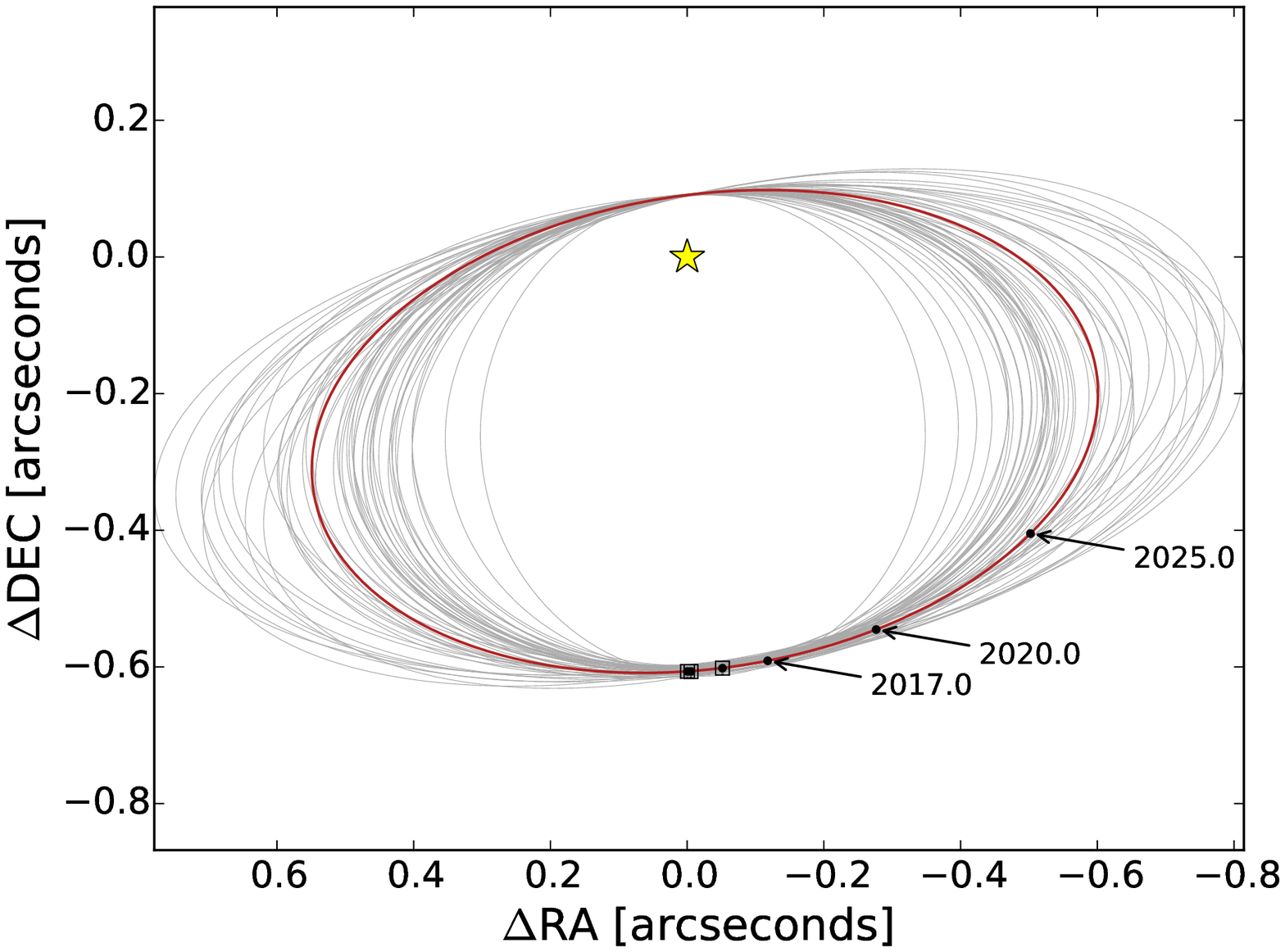} 
\caption{The orbit of HD 4747~B as projected onto the sky (astrometry displayed as black squares including uncertainties) showing our best model fit (red) along with 50 randomly drawn parameter sets from the MCMC chain (black). Epochs from the years 2017, 2020, and 2025 are displayed for reference to show how the orbit can be constrained with future imaging measurements.}
\label{fig:orbit}
\end{figure}

\begin{figure*}[!t]
\includegraphics[height=3.74in]{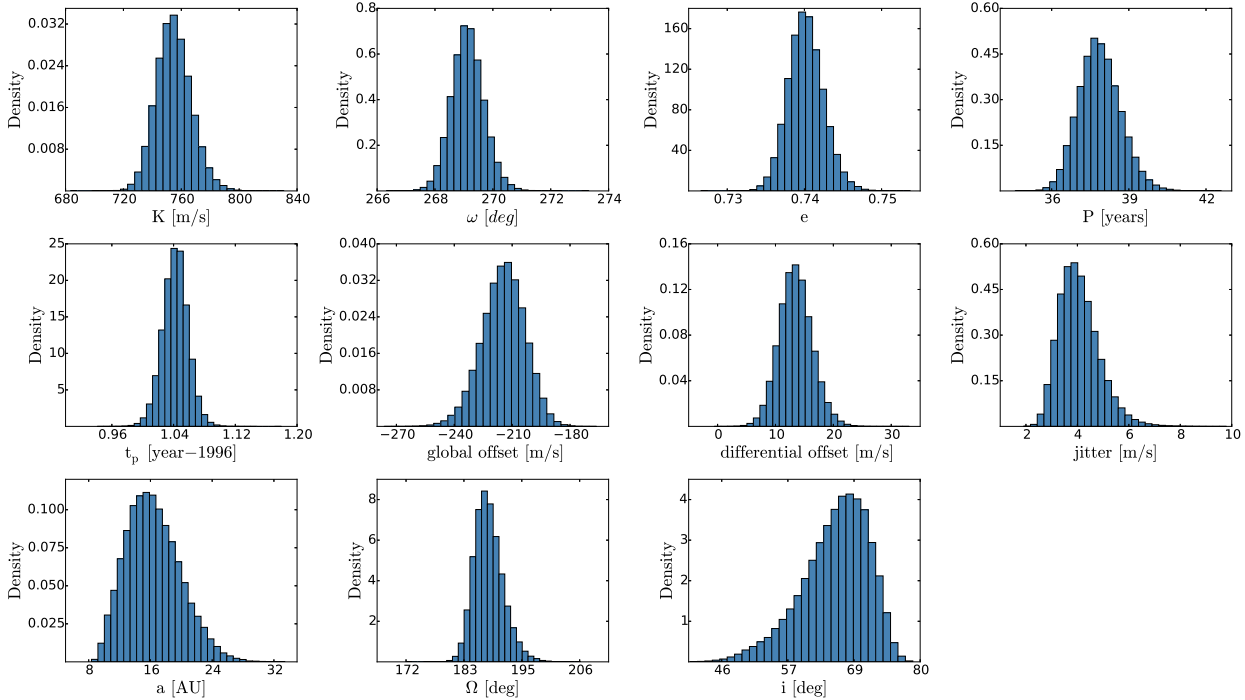} 
\caption{Posterior distribution for parameters involved in the HD~4747~B orbital fitting procedure.}
\label{fig:posterior}
\end{figure*}

Fig.~\ref{fig:posterior} shows the posterior distributions used to derive the uncertainty for each orbital parameter in the analysis. In addition to the self-consistent modeling of RV data with astrometry, we also analyzed the RV data alone with a separate MCMC run. Perhaps not surprisingly, we find that the RV data dominates the fit for most parameters including those shared in common between Doppler measurements and imaging ($P$, $e$, $\omega$, $t_p$). Fig.~\ref{fig:orbit} shows the orbit of HD~4747~B as projected onto the sky along with example orbital fits from our MCMC statistical procedure. 

\begin{figure*}[!t]
\includegraphics[height=6.5in]{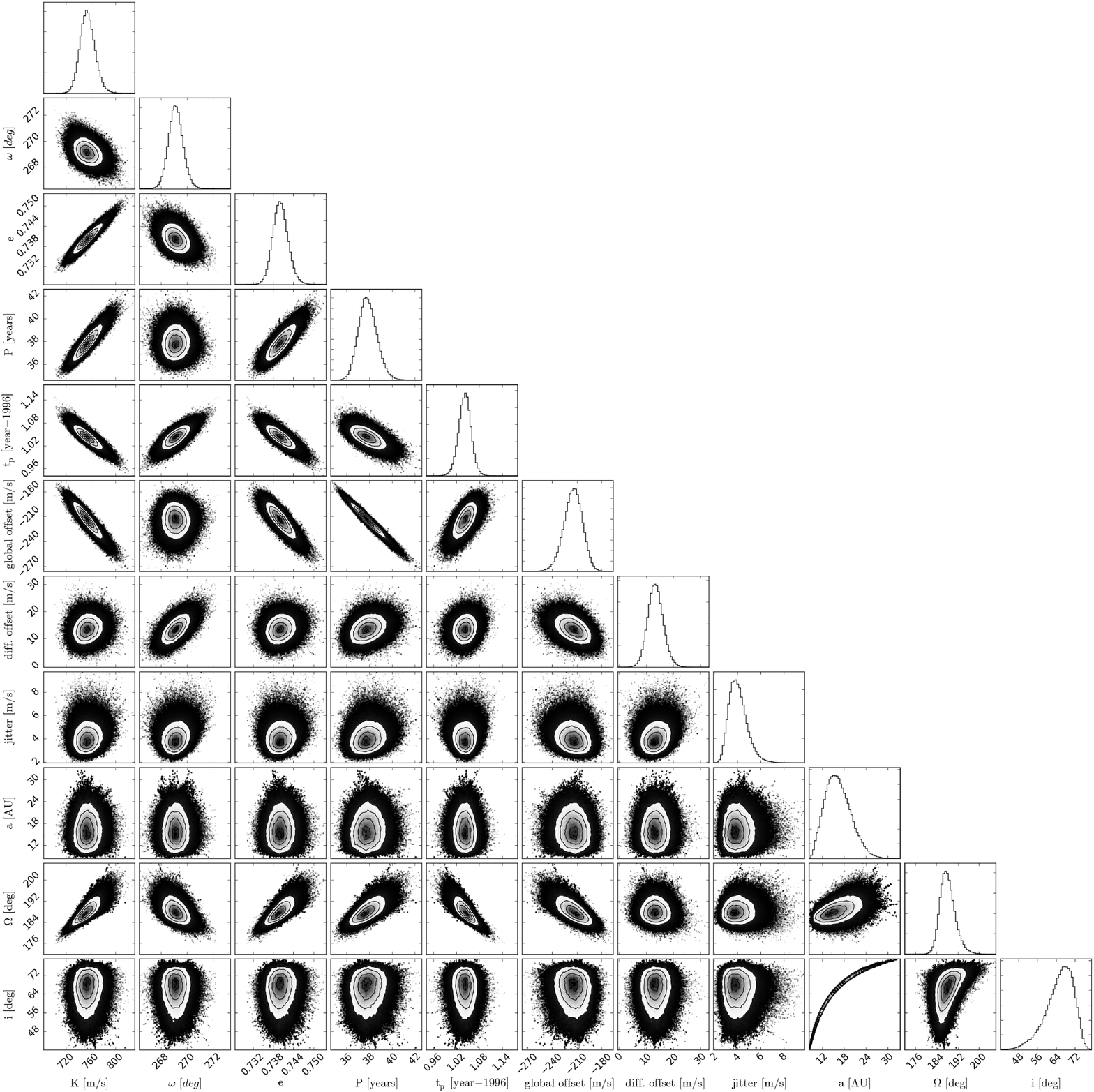} 
\caption{Matrix showing correlations between orbit fitting parameters. Notice that semi-major axis and inclination are strongly degenerate. This in turn affects the total mass estimate for the HD~4747 system. See text for discussion.}
\label{fig:triangle}
\end{figure*} 

We take several approaches to estimating the companion mass. First, we calculate $m \sin{i}$ from the velocity semi-amplitude and orbital eccentricity assuming a host star mass of $M_*=0.82\pm0.04M_{\odot}$ as estimated from an isochrone analysis (Table~\ref{tab:starprops}). Specifically, from the mass function $f(m)=\frac{(m \sin i)^3}{(M_*+m)^2}=K^3 P \frac{(1-e^2)^{3/2}}{2 \pi G}=3.7\pm0.2 \times 10^{26}$ kg, we find $m \sin{i}=55.3\pm1.9 M_{\rm Jup}$.

Next, we place a lower limit on the companion mass using the instantaneous acceleration (slope) of the RV time series evaluated during the direct imaging discovery \citep{torres_99}. Using this method, the mass of the star is automatically encoded in the measured Doppler trend itself thus eliminating any systematic errors resulting from model-dependent estimates of the host star mass. Using the relationship between distance, angular separation, and RV acceleration from \citet{liu_02} (see also \citet{rodigas_16}), the minimum dynamical mass of HD~4747~B is $m_{\rm trend} \geq 57.3\pm2.0 M_{\rm Jup}$. Comparing this value with the $m \sin{i}$ value and mass estimated from thermal evolutionary models, we find that the companion physical separation and brightness are indeed consistent with producing the measured RV acceleration of $\dot{v}=-30.7\pm0.6$ m/s/yr. 

Finally, we calculate the companion mass directly using the results of our MCMC analysis. Breaking the $m \sin{i}$ degeneracy with astrometry, we find $m=55.3\pm1.9M_{\rm Jup}/ \sin\left({66.8^{+4.7}_{-6.4} \; \mbox{deg}}\right)=60.2\pm3.3M_{\rm Jup}$. This result is consistent with both the companion minimum dynamical mass found from the RV acceleration and also the predicted mass from thermal evolutionary models. Table~\ref{tab:mass} summarizes the myriad ways for constraining the companion mass using this rich data set. 

To determine the system total mass, we use Kepler's equation again with our best-fit semi-major axis and orbital period. 

\begin{equation}
M_*+m=\frac{a^3}{p^2}
\end{equation}

Propagating the uncertainties from our (asymmetric) posterior distributions we find $M_*+m=2.9^{+2.6}_{-1.4}M_{\odot}$. This value is clearly much larger than the actual system mass since we already know the host star mass to 5\% from spectroscopic modeling ($M_*=0.82\pm0.04M_{\odot}$). The orbital period is known to 2\% (68\% confidence) but the semi-major axis, which enters as the third-power, is still highly uncertain as the result of having only three astrometric measurements. Inverting the problem to instead estimate $a$ from $M_*$ (from high resolution spectroscopy) and $m$ (estimated coarsely from thermal models), the semimajor axis has a most-probable value of $a\approx10.7$ AU; thus existing astrometry from a limited time baseline may be over-estimating $a$ by as much as $\sim50\%$. This result in turn affects the orbital inclination which influences the total mass estimate. As shown in Fig.~\ref{fig:triangle}, semi-major axis and orbital inclination are strongly correlated. Clearly more astrometry is needed to improve our understanding of HD~4747's orbital parameters.  


\begin{table}[!ht]
\centerline{
\begin{tabular}{lc}
\hline
\hline
\multicolumn{2}{c}{HD~4747~B}     \\
\hline
\hline
$m \sin i$ [$M_{\rm Jup}$]                  &    $55.3\pm1.9$  \\ 
$m_{\rm trend}$ [$M_{\rm Jup}$]        &    $\geq57.3\pm2.0$  \\
$m_{\rm 3d-orbit}$ [$M_{\rm Jup}$]    &    $60.2\pm3.3$  \\
$m_{\rm model}$ [$M_{\rm Jup}$]      &    $72^{+3}_{-13}$ \\ 
\hline
\hline
\end{tabular}}
\caption{Various mass estimates of HD~4747~B. See text for discussion.}
\label{tab:mass}
\end{table} 

\section{SUMMARY} 
Few substellar objects have been detected in the brown dwarf ``desert," let alone imaged directly. We have discovered a rare benchmark object orbiting the nearby ($d=18.69\pm0.19$ pc), bright ($V=7.16$), solar-type (G9V) star HD~4747 that not only lends itself to precise dynamical studies but also has an age ($3.3^{+2.3}_{-1.9}$ Gyr) and metallicity ([Fe/H]=$-0.22\pm0.04$) that are known independently from the light that the brown dwarf emits. Only several other non-transiting brown dwarfs provide all three pieces of information simultaneously \citep{liu_02,potter_02,montet_15}. Following 18 years of precise RV measurements, we were able to estimate the approximate angular separation of HD~4747's orbiting companion as previous observations of the star had narrowed down the parameter space of possible orbital orientations and mass. Our NIRC2 observations in two separate filters across three distinct epochs confirm the companion as co-moving and also reveal orbital motion in a counter-clockwise direction. 

We find consistency between HD~4747~B's intrinsic brightness ($M_{K_s}=13.00\pm0.13$), infrared color ($K_s-L'=1.34\pm0.46$), and minimum mass of $m_{\rm trend}\geq57.3\pm2.0$ based on the instantaneous RV acceleration and measured angular separation. Placing the companion on a color-magnitude diagram suggests that it is a late L-dwarf possibly near the L/T transition. Follow-up spectroscopy is warranted to refine the companion spectral-type, effective temperature, and assess surface gravity. 

Jointly modeling the RV data and astrometry from imaging, we were are able to calculate the companion mass and orbit using standard statistical methods (MCMC). We find that the companion has a remarkably large orbital eccentricity of $e=0.740\pm0.002$ hinting at an interesting dynamical history. We find a preliminary dynamical mass of HD~4747~B of $m_{\rm 3d-orbit}=60.2\pm3.3M_{\rm Jup}$. This result is consistent with thermal evolutionary models which predict $m_{\rm model}=72^{+3}_{-13}M_{\rm Jup}$ given the stellar gyro-chronological age. We caution however that this value is only preliminary given the limited astrometry time baseline and subject to a wide and asymmetric posterior distribution in the system semi-major axis and orbital inclination. As such, HD~4747~B will immediately benefit from follow-up high-contrast imaging measurements which should be able to pin-down the companion mass to only several percent. Instruments like the Gemini Planet Imager \citep{macintosh_15} and SPHERE \citep{vigan_16} could provide both spectroscopy and astrometry simultaneously aiding significantly in the study of HD~4747~B ($\delta=-23^{\circ}$).

\section{ACKNOWLEDGEMENTS}
We thank the many California Planet Search observers for help over the years securing precise RV measurements that ultimately led to the direct imaging discovery of HD~4747~B. Chris Matthews estimated the $L'$ apparent magnitude of HD~4747~A by fitting an SED to its flux measured at visible and near-infrared wavelengths. This work was supported by a NASA Keck PI Data Award, administered by the NASA Exoplanet Science Institute. Data presented herein were obtained at the W. M. Keck Observatory from telescope time allocated to the National Aeronautics and Space Administration through the agencyÕs scientific partnership with the California Institute of Technology and the University of California. The Observatory was made possible by the generous financial support of the W. M. Keck Foundation. B.T.M. is supported by the National Science Foundation Graduate Research Fellowship under Grant No. DGE-1144469. The TRENDS high-contrast imaging program is supported in part by NASA Origins grant NNX13AB03G and PI Crepp's NASA Early Career Fellowship. We are also grateful for the vision and support of the Potenziani family. 


\begin{table*}[!ht]
\centerline{
\begin{tabular}{ccc}
\hline
\hline
HJD-2,440,000  &   RV [m~s$^{-1}$]    & Uncertainty [m~s$^{-1}$]   \\
\hline
\hline
10366.975	&	-541.03	&	1.23	\\
10418.829	&	-400.53	&	2.06	\\
10462.752	&	-240.94	&	1.10	\\
10690.104	&	372.92	&	1.17	\\
10784.771	&	462.23	&	1.33	\\
10806.727	&	485.19	&	1.03	\\
10837.717	&	493.95	&	1.17	\\
10838.701	&	497.51	&	1.12	\\
10839.711	&	501.87	&	1.12	\\
11010.120	&	533.55	&	1.35	\\
11012.048	&	531.64	&	1.28	\\
11014.102	&	536.29	&	1.60	\\
11050.991	&	521.54	&	4.68	\\
11170.748	&	507.57	&	1.40	\\
11368.085	&	476.78	&	1.31	\\
11410.027	&	461.47	&	5.23	\\
11543.744	&	443.11	&	1.52	\\
11550.729	&	432.86	&	1.47	\\
11756.051	&	393.29	&	2.13	\\
11899.784	&	364.57	&	1.63	\\
12098.126	&	329.86	&	1.57	\\
12489.042	&	266.87	&	1.50	\\
12572.812	&	250.35	&	1.95	\\
12987.728	&	194.66	&	2.38	\\
12988.686	&	184.69	&	1.94	\\
13238.956	&	168.92	&	1.34	\\
13303.903	&	167.69	&	1.12	\\
13339.765	&	160.69	&	1.19	\\
13724.774	&	113.02	&	1.03	\\
13724.776	&	116.05	&	1.12	\\
14717.955	&	9.8418	&	1.23	\\
14719.008	&	12.417	&	1.15	\\
14720.010	&	15.718	&	1.23	\\
14720.970	&	15.336	&	1.13	\\
14722.901	&	14.786	&	1.27	\\
14723.970	&	14.225	&	1.22	\\
14724.974	&	11.688	&	1.20	\\
14725.883	&	13.827	&	1.13	\\
14727.005	&	19.502	&	1.23	\\
14727.952	&	17.496	&	1.13	\\
 \hline
\end{tabular}}
\caption{Precise RV measurements of the star HD~4747.}
\label{tab:rvtable}
\end{table*}

\begin{table*}[!ht]
\centerline{
\begin{tabular}{ccc}
\hline
\hline
HJD-2,440,000  &   RV [m~s$^{-1}$]    & Uncertainty [m~s$^{-1}$]   \\
\hline
\hline
15015.120	&	-6.36		&	1.41	\\
15016.124	&	-4.09		&	1.38	\\
15017.124	&	-5.66		&	1.21	\\
15049.025	&	-19.75	 &	1.24	\\
15077.084	&	-16.14	 &	1.28	\\
15078.094	&	-13.03 	&	1.26	\\
15134.964	&	-16.89 	&	1.36	\\
15198.754	&	-28.85	&	1.20	\\
15426.084	&	-43.59	&	1.25	\\
15522.879	&	-51.20	&	1.30	\\
15807.047	&	-76.97	&	1.30	\\
16149.050	&	-104.29	&	1.28	\\
16319.701	&	-110.20	&	1.28	\\
16327.708	&	-113.32	&	1.49	\\
16508.145	&	-123.24	&	1.33	\\
16913.034	&	-170.85	&	1.13	 \\
 \hline
\end{tabular}}
\caption{Table~\ref{tab:rvtable} continued.}
\end{table*}

\end{document}